\documentclass[superscriptaddress,aps, prd, reprint, nofootinbib]{revtex4-2}

\usepackage{graphicx}
\usepackage{amsmath,amssymb,mathtools}
\usepackage{dcolumn}
\usepackage{bm}
\usepackage{xcolor}
\usepackage{soul}
\usepackage{braket}
\usepackage[separate-uncertainty=true,multi-part-units=single]{siunitx}

\newcommand{\dd}{\mathop{\mathrm{d}\!}{}}
\newcommand{\lab}[1]{{\mathrm{#1}}}

\newcommand{\minus}{{\protect \scalebox {0.75}[1.0]{$-$}}}

\newcommand{\floq}[1]{{\scalebox{0.65}{$(#1)$}}}

\renewcommand{\vec}[1]{\boldsymbol{\mathbf{#1}}}

\renewcommand{\Im}{\operatorname{Im}}

\DeclarePairedDelimiter{\abs}{\lvert}{\rvert}

\definecolor{blue2}{cmyk}{1, 0.1, 0.1, 0.1}

\usepackage{colortbl}
\definecolor{lightgreen}{cmyk}{0.2, 0, 0.2, 0.2}
\definecolor{lightgray2}{cmyk}{0.1,0.1,0,0.1}
\definecolor{Red2}{RGB}{214, 39, 40}
\definecolor{Blue2}{RGB} {31, 119, 180}
\definecolor{Orange2}{RGB}{255, 127, 14}
\definecolor{Green2}{RGB}{44, 160, 44}
\definecolor{greyish2}{rgb}{.96,.96,.96}

\definecolor{pyBlue}{RGB}{31, 119, 180}
\definecolor{pyRed}{RGB}{214, 39, 40}
\definecolor{pyGreen}{RGB}{44, 160, 44}
\definecolor{pyBlue2}{RGB}{0, 111, 237}
\definecolor{pyRed2}{RGB}{224, 52, 36}
\definecolor{Mathematica1}{rgb}{0.368417, 0.506779, 0.709798}
\definecolor{Mathematica2}{rgb}{0.880722, 0.611041, 0.142051}

\def\beq{\begin{equation}}
\def\eeq{\end{equation}}

\begin{document}

\title{Legacy of Boson Clouds on Black Hole Binaries}

\author{Giovanni Maria Tomaselli}
\affiliation{Gravitation Astroparticle Physics Amsterdam (GRAPPA),
University of Amsterdam, Amsterdam, 1098 XH, Netherlands}

\author{Thomas F.M.~Spieksma}
\affiliation{Niels Bohr International Academy, Niels Bohr Institute, 
Blegdamsvej 17, 2100 Copenhagen, Denmark}

\author{Gianfranco Bertone}
\affiliation{Gravitation Astroparticle Physics Amsterdam (GRAPPA),
University of Amsterdam, Amsterdam, 1098 XH, Netherlands}

\begin{abstract}

Superradiant clouds of ultralight bosons can leave an imprint on the gravitational waveform of black hole binaries through ``ionization'' and ``resonances.'' We study the sequence of resonances as the binary evolves, and show that there are only two possible outcomes, each with a distinct imprint on the waveform. If the cloud and the binary are nearly counter-rotating, then the cloud survives in its original state until it enters the sensitivity band of future gravitational wave detectors, such as LISA. In all other cases, resonances destroy the cloud, while driving the binary to co-rotate with it and its eccentricity close to a fixed point. This opens up the possibility of inferring the existence of a new boson from the statistical analysis of a population of black hole binaries.

\end{abstract}

\maketitle
\paragraph*{\bf  Introduction.}
Gravitational waves (GWs) will enable detailed measurements of the environments of compact objects, especially for intermediate-to-extreme mass ratio inspirals \cite{Barausse:2014tra,Cole:2022yzw}. These systems spend millions of cycles in the millihertz regime, allowing environmental effects to build up throughout the inspiral. Future GW detectors, such as LISA \cite{LISA:2017pwj,Baker:2019nia}, will probe this frequency range; hence, a precise modelling of the dynamics between the binary and the environment is essential. 

Before the binary's frequency enters the detector's band, the companion object may have already impacted the environment through its gravitational perturbation. Understanding the ``history'' of the system is, thus, required to study effects later in the inspiral, such as dynamical friction. A type of environment that has recently attracted significant attention are clouds made of hypothetical ultralight bosons surrounding black holes (BHs). It has been found that the cloud-BH system, known as a ``gravitational atom,'' exhibits a rich phenomenology when part of a binary system \cite{Zhang:2018kib,Baumann:2018vus,Zhang:2019eid,Baumann:2019ztm,Baumann:2021fkf,Baumann:2022pkl,Tomaselli:2023ysb,Tomaselli:2024bdd,Boskovic:2024fga,Brito:2023pyl,Duque:2023cac}, and, thus, forms an interesting candidate for precision tests of fundamental physics with GWs.

Boson clouds have a natural way of forming through \emph{superradiance} \cite{ZelDovich1971,ZelDovich1972,Starobinsky:1973aij,Brito:2015oca}. This is a purely gravitational process where rotating BHs shed a significant amount of their mass and angular momentum to a boson field, and can be excited without the need of any prior abundance. The efficiency of this process depends on the gravitational coupling\footnote{We use natural units $G=\hbar=c=1$.} $\alpha=\mu M$, where $\mu$ is the boson mass and $M$ the BH mass, and is maximum when $\alpha\sim\mathcal O(0.1)$. For astrophysical BHs, this condition implies that the boson has to be ultralight, i.e., in the mass range $\mathcal{O}(10^{-20}\!-\!10^{-10})\,\text{eV}$. These types of particles are predicted from different areas of physics, for example in the context of the strong CP problem (the QCD axion \cite{Weinberg:1977ma,Wilczek:1977pj,Peccei:1977hh}), the string axiverse \cite{Arvanitaki:2009fg,Svrcek:2006yi} and as possible candidates to resolve the dark matter problem \cite{Bergstrom:2009ib,Marsh:2015xka,Hui:2016ltb,Ferreira:2020fam}. 

The cloud-binary interaction involves \emph{ionization} (or ``dynamical friction'') \cite{Baumann:2021fkf,Baumann:2022pkl,Tomaselli:2023ysb} and \emph{resonances} \cite{Baumann:2018vus,Baumann:2019ztm,Tomaselli:2024bdd}. While ionization occurs ``late'' in the inspiral, resonances can affect the system when the binary is well beyond the cloud's radius, making them crucial for understanding the history of the system. In this \emph{Letter} and its companion \cite{Tomaselli:2024bdd}, we study this resonant behavior on orbits with generic eccentricity and inclination and uncover new phenomena that can cause a resonance to ``break'' before its completion. We are able to (i) narrow down \emph{direct} observational signatures from the cloud, when it survives all resonances, and (ii) discover new \emph{indirect} observational signatures, when the cloud is destroyed early on, thereby leaving a legacy on the binary through its eccentricity and inclination.

\vskip 4pt
\paragraph*{\bf Setup.} We work in the frame of the larger BH, assumed to host the cloud, with $\vec{r} = \{r, \theta, \phi\}$. We specialize to scalar fields and work in the nonrelativistic limit, where the Klein-Gordon equation reduces to the Schrödinger equation and is solved by the familiar hydrogenic eigenfunctions, $\psi_{n\ell m} = R_{n \ell}(r)Y_{\ell m}(\theta, \phi) e^{-i(\omega_{n\ell m} - \mu) t}$. Here, $R_{n\ell}$ are the hydrogenic radial functions, $Y_{\ell m}$ the scalar spherical harmonics, and $n$, $\ell$, and $m$ the usual quantum numbers. The energy of each eigenstate is given by 
\beq
\epsilon_{n\ell m}=\mu\biggl(1-\frac{\alpha^{2}}{2 n^{2}}-F_{n\ell}\alpha^4+\frac{h_\ell}{n^3}\tilde{a} m \alpha^{5}+\mathcal{O}(\alpha^{6})\biggr)\,,
\label{eq:eigenenergy}
\eeq
where $\tilde a$ is the spin of the BH and the coefficients $F_{n\ell}$ and $h_\ell$ can be found in \cite{Baumann:2019eav}. The energy difference between two states is referred to as \emph{Bohr} ($\Delta n \neq 0$), \emph{fine} ($\Delta n = 0$, $\Delta \ell \neq 0$) or \emph{hyperfine} ($\Delta n =\Delta \ell = 0$, $\Delta m \neq 0$), depending on the leading power of $\alpha$.

The companion object has mass $M_*$, position $\vec{R}_{*} = \{R_*, \theta_*, \varphi_*\}$ and its Newtonian gravitational perturbation can be expanded in multipoles as
\beq
\label{eqn:V_star}
V_*(t,\vec r)=-\sum_{\ell_*, m_*}\!\frac{4\pi q\alpha}{2\ell_*+1}\frac{r^{\ell_*}_{\scalebox{0.60}{$\mathrm{<}$}}}{r^{\ell_*+1}_{\scalebox{0.60}{$\mathrm{>}$}}}Y_{\ell_*m_*}(\theta_*,\varphi_*)Y_{\ell_*m_*}^*(\theta,\phi)\,
\eeq
for $\ell_* \neq 1$, where $r_{\scalebox{0.60}{$\mathrm{>}$}}$ $(r_{\scalebox{0.60}{$\mathrm{<}$}})$ indicates the larger (smaller) of $R_*$ and $r$ and $q=M_*/M$ is the mass ratio. The dipole contribution $\ell_* = 1$ has a slightly different expression and can be found in \cite{Tomaselli:2023ysb}.

\vskip 4pt
\paragraph*{\bf Nonlinear resonances.} The overlap between two different eigenstates $\ket{a}$ and $\ket{b}$ induced by the perturbation \eqref{eqn:V_star} is composed of terms that oscillate with an integer multiple $g$ of the orbital frequency $\Omega=\dot\varphi_*$, $\braket{a|V_*|b}\sim\eta^\floq{g}e^{ig\varphi_*}$. Energy losses such as GW radiation and the cloud's ionization induce a slow frequency chirp, that can be linearized around a reference point $\Omega_0$ as $\Omega=\gamma t+\Omega_0$, where $\gamma$ quantifies the speed of the chirp. Restricting our attention to two eigenstates, the Schrödinger equation can be written in a dimensionless form as
\beq
\frac\dd{\dd\tau}\!\begin{pmatrix}c_a\\ c_b\end{pmatrix}=-i\begin{pmatrix}\omega/2 & \sqrt{Z}\\ \sqrt{Z} & -\omega/2-i\Gamma\end{pmatrix}\begin{pmatrix}c_a\\ c_b\end{pmatrix}\,,
\label{eqn:2-state}
\eeq
where $\tau=\sqrt{\abs{g}\gamma}\,t$ and $c_j=\braket{j|\psi}$, with $j=a,b$. The parameter $\Gamma$ quantifies the decay of state $\ket{b}$ \cite{Baumann:2019eav}, the strength $\eta^\floq{g}$ of the perturbation \eqref{eqn:V_star} defines the \emph{Landau-Zener} (LZ) \emph{parameter} $Z=(\eta^\floq{g})^2/(\abs{g}\gamma)$, and the dimensionless frequency is given by $\omega=(\Omega-\Omega_0)/\sqrt{\gamma/\abs{g}}$, where $\Omega_0=(\epsilon_b-\epsilon_a)/g$ is the \emph{resonance frequency}. As long as $\omega$ increases linearly, the long-time behavior of the linear system \eqref{eqn:2-state} follows the well-known LZ formula \cite{zener1932non,landau1932theorie}: if initially only state $\ket{a}$ is populated, its final population is given by $|c_a|^2=e^{-2\pi Z}$.

However, the transition between states of the cloud induces a backreaction on the orbit. From the energy balance of the cloud-binary system, the frequency is found to increase as
\beq
\omega=\tau-B\abs{c_b}^2\,,
\label{eqn:omega-B}
\eeq
where the dimensionless parameter $B$ quantifies the strength of the backreaction and is given by
\beq
B=\frac{3M_{\rm c}}M\frac{\Omega_0^{4/3}(1+q)^{1/3}M^{1/3}}{q\alpha\sqrt{\gamma/\abs{g}}}(-g)\,.
\eeq
The system \eqref{eqn:2-state}--\eqref{eqn:omega-B} is now nonlinear, and behaves differently from the LZ solution. Two cases can be distinguished based on the sign of $B$.

If $B>0$, the backreaction slows down the chirp. The equations then enjoy a positive-feedback mechanism: the stronger the backreaction, the more population is transferred between the states, which, in turn, makes the backreaction stronger, and so on. Hence, there is a critical threshold above which this process becomes self-sustaining and the system enters a distinct phase of a \emph{floating orbit} during which $\Omega$ stays approximately constant. The value of the tipping point can be estimated perturbatively for $2\pi Z\ll1$: ignoring the backreaction, $\abs{c_b}^2$ would reach the value $1-e^{-2\pi z}\approx2\pi Z$ in a time interval $\tau\approx1$; the backreaction term in \eqref{eqn:omega-B} must then become significant when
\beq
2\pi ZB\gtrsim1\qquad\text{(float starts).}
\label{eqn:threshold}
\eeq
Careful numerical study of the nonlinear set of eqs.~\eqref{eqn:2-state}--\eqref{eqn:omega-B} confirms this formula to an accuracy better than $6\%$.

While \eqref{eqn:threshold} determines whether the float starts, a complete transfer from $\ket{a}$ to $\ket{b}$ is realized only for small enough $\Gamma$. In particular, it can be shown analytically \cite{Tomaselli:2024bdd} that, for $\Gamma B\gg1$, the float \emph{breaks} abruptly when the population left in the initial state is
\beq
\abs{c_a}^2\approx\frac{\Gamma}{2ZB}\qquad\text{(float breaks),}
\label{eqn:Gamma-breaking}
\eeq
while $\abs{c_b}^2$ remains at negligible values throughout and after the resonance.\footnote{When formula \eqref{eqn:Gamma-breaking} returns $\abs{c_a}^2>1$, the resonance breaks as soon as it starts, effectively behaving as if \eqref{eqn:threshold} was not satisfied.} Together, \eqref{eqn:threshold} and \eqref{eqn:Gamma-breaking} characterize the nonlinear behavior of floating resonances.

For $B<0$ instead, the backreaction speeds up the chirp and reduces the population transferred. For small $q$, we have $2\pi Z\ll1$, and the final population in state $\ket{b}$ is found to be
\beq
\abs{c_b}^2\approx3.7\,\biggl(\frac{Z}{B^2}\biggr)^{1/3}\,,
\label{eqn:scaling-cb2}
\eeq
which is valid for $B\ll-1/Z$. While the numerical coefficient in \eqref{eqn:scaling-cb2} is determined with a numerical fit, the dependence on $Z$ and $B$ can be justified by keeping only the second term in \eqref{eqn:omega-B}, which dominates for large negative $B$, and then looking for a stationary solution of \eqref{eqn:2-state} with small $c_b$. The backreaction now produces a \emph{sinking orbit}, during which the frequency ``jumps'' ahead of its unbackreacted value in a short interval of time.

\vskip 4pt
\paragraph*{\bf Eccentricity and inclination.} Any general conclusion about the role of resonances in the evolution of the cloud-binary system must take into account generic orbital configurations, with nonzero eccentricity $\varepsilon$ and inclination $\beta$. While on equatorial circular orbits resonances are excited only at $\Omega_0=\Delta\epsilon/\Delta m$ (that is, only for the \emph{main tone} $g=\Delta m$), on generic orbits the gravitational perturbation includes \emph{overtones} $g\ne\Delta m$, allowing a resonance between two states to be excited at multiple points in the inspiral. Furthermore, a much larger number of resonances is now possible, virtually between any two given states.

Eccentricity and inclination also modulate the chirp rate $\gamma$ and the strength $\eta^\floq{g}$ of the perturbation. For example, on circular orbits, the LZ parameter of the most interesting hyperfine and fine resonances depends on the inclination as
\beq
Z\propto\sin^{2(\Delta m-g)}(\beta/2)\cos^{-2(\Delta m+g)}(\beta/2)\,,
\label{eqn:Z-beta}
\eeq
meaning that these resonances all become weak close to counter-rotating orbits ($\beta=\pi$).

Most importantly, however, resonances severely impact the orbital parameters with their backreaction. First of all, eq.~\eqref{eqn:omega-B} is modified to
\beq
\frac{\dd\omega}{\dd\tau}=f(\varepsilon)-B\frac{\dd\abs{c_b}^2}{\dd\tau}\,,\quad\;\ f(\varepsilon)=\frac{1+\frac{73}{24}\varepsilon^2+\frac{37}{96}\varepsilon^4}{(1-\varepsilon^2)^{7/2}}\,,
\eeq
where $f(\varepsilon)$ quantifies the eccentricity dependence of GW energy losses \cite{Peters:1963ux,Peters:1964zz}. Then, other dynamical equations for $\varepsilon$ and $\beta$ arise from the conservation of angular momentum. On floating orbits, which are particularly efficient in changing the orbital parameters due to their long duration, these read
\begin{align}
\label{eqn:eccentricity-evolution-floating}
C\frac\dd{\dd\tau}\sqrt{1-\varepsilon^2}&=\frac{\Delta m}gf(\varepsilon)\cos\beta-h(\varepsilon)\,,\\
\label{eqn:inclination-evolution-floating}
C\sqrt{1-\varepsilon^2}\frac{\dd\beta}{\dd\tau}&=-\frac{\Delta m}gf(\varepsilon)\sin\beta\,,
\end{align}
where $C=3\Omega_0/\sqrt{\gamma/\abs{g}}$ and $h(\varepsilon)=(1+7\varepsilon^2/8)/(1-\varepsilon^2)^2$.

\begin{figure}
\centering
\includegraphics[width=\columnwidth]{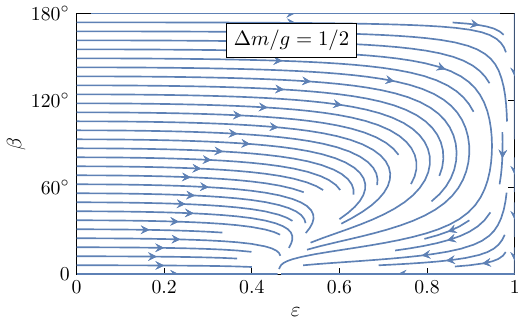}
\caption{Evolution of eccentricity $\varepsilon$ and inclination $\beta$ assuming that the system on a floating orbit. For illustrative purposes, we chose $\Delta m/g=1/2$, in which case there is a fixed point at $(\varepsilon,\beta)\approx(0.46,0)$.}
\label{fig:streamplot}
\end{figure}

The flow defined by \eqref{eqn:eccentricity-evolution-floating} and \eqref{eqn:inclination-evolution-floating} is shown in Fig.~\ref{fig:streamplot} for an example value of $\Delta m/g$. Each overtone forces the parameters to evolve toward a fixed point where the binary is co-rotating ($\beta=0$) and the eccentricity depends on $\Delta m/g$, for example
\beq
\begin{array}{cccccc}
\qquad\,\varepsilon\approx{} & \ldots, & 0.65, & 0.58, & 0.46, & 0\,,\\
\Delta m/g={} & \ldots, & 1/4, & 1/3, & 1/2, & 1\,.
\end{array}
\label{eqn:fixedpoints}
\eeq
If the resonance does not break before completion, it lasts for a time $\Delta\tau=B$, and its ``distance'' from the fixed point in the $(\varepsilon,\beta)$ plane roughly reduces by a factor $e^D$, where $D\equiv B/C\propto M_\lab{c}q^{-1}$. Unequal mass binaries, thus, approach the fixed point more than equal mass ones.

Finally, we note that an increase in $\varepsilon$ or decrease in $Z$ during the float can break the resonance, similar to the effect of the decay of state $\ket{b}$, while variations in the opposite direction can prevent the break. For the main tones ($g=\Delta m$), however, these effects turn out to not be relevant \cite{Tomaselli:2024bdd}.

\vskip 4pt
\paragraph*{\bf Resonant history.} Predicting the nonlinear behavior of resonances on generic orbits allows us to determine the evolution of the cloud-binary system. This includes answering the pressing question of whether the cloud is still present (and in which state) when the binary is close enough to give direct observational signatures, through Bohr resonances \cite{Baumann:2018vus,Baumann:2019ztm} and ionization \cite{Baumann:2021fkf,Baumann:2022pkl,Tomaselli:2023ysb}.

The answer is found studying hyperfine and fine resonances. All of them are floating, since superradiance populates the most energetic state for a given $n$, such as $\ket{211}$ or $\ket{322}$, see \eqref{eq:eigenenergy}. Furthermore, these resonances all involve a \emph{decaying} final state, i.e., with $\Im(\omega)<0$. This decay is always much faster than duration of the float $\Delta t_{\rm float}$, especially for small mass ratios, where the two timescales are separated by many orders of magnitude; see Fig.~10 in \cite{Tomaselli:2024bdd}.

We can, thus, already conclude that none of the early resonances is able to change the state of the cloud: either it survives in its original state $\ket{a}\equiv\ket{n_a\ell_am_a}$, or it is destroyed. It remains to be answered under which conditions either possibility occurs. The overtones generated by a nonzero eccentricity are generally weaker than the resonances encountered on circular orbits. So, if the cloud safely gets past the strongest hyperfine or fine resonance on circular orbits, it will still be present when the binary eventually enters the Bohr region.

\begin{figure}
\centering
\includegraphics[width=0.95\columnwidth,trim={0pt 6pt 0pt 0}]{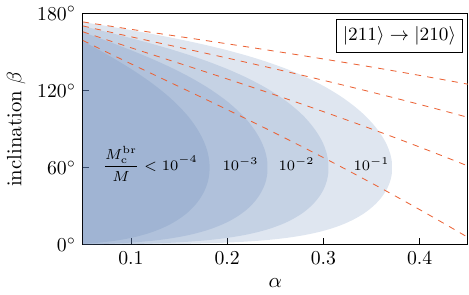}
\caption{Mass of the cloud $M_{\rm c}^{\rm br}$ at resonance breaking, for $\ket{211}\overset{g=-2}\longrightarrow\ket{210}$ on circular orbits, as function of $\alpha$ and $\beta$. The dashed red lines correspond to the analytical approximation \eqref{eqn:chi_1}.}
\label{fig:breaking} 
\end{figure}

Two conditions need to be satisfied for a floating resonance to destroy the cloud. First, the float must \emph{start}; i.e., inequality \eqref{eqn:threshold} must hold. Second, the float must not \emph{break} before the cloud is destroyed. The mass left in the original state when the resonance breaks is given in \eqref{eqn:Gamma-breaking} and is shown in Fig.~\ref{fig:breaking} for the example resonance $\ket{211}\to\ket{210}$. Both conditions fail when the resonance is weak, for example, when $Z\to0$. From \eqref{eqn:Z-beta}, it follows that this is necessarily the case in a certain interval of inclinations that neighbors the counter-rotating configuration, say $\pi-\chi_i<\beta\le\pi$, where $i=1$ ($i=2$) for hyperfine (fine) resonances. Not breaking the resonance is often a stronger condition than starting it, and it is, thus, the one that determines $\chi_i$. Analytical approximations for $\alpha\ll1$ give
\beq
\chi_1\approx\Theta\times\biggl(\frac{M_{\rm c}^{\rm br}/M}{10^{-2}}\biggr)^{-1/6}\biggl(\frac\alpha{0.2}\biggr)^{(3+4\ell_a)/6}\biggl(\frac{\tilde a}{0.5}\biggr)^{-5/18}\,,
\label{eqn:chi_1}
\eeq
where $\Theta=\SI{38}{\degree}$ for $\ket{211}$ and $\Theta=\SI{4.8}{\degree}$ for $\ket{322}$, while $M_{\rm c}^{\rm br}$ is the cloud's mass at resonance breaking. Fine resonances, instead, give
\beq
\chi_2\approx\SI{9}{\degree}\biggl(\frac{M_{\rm c}^{\rm br}/M}{10^{-2}}\biggr)^{-1/4}\biggl(\frac\alpha{0.2}\biggr)^{3/2}\,.
\label{eqn:322_chi_2}
\eeq
for $\ket{322}$, while they are never able to destroy a cloud in the $\ket{211}$ state.

Although resonances with $g\ne\Delta m$ also fail to destroy the cloud in other angular intervals, such as near-co-rotating binaries ($\beta=0)$, the one around $\beta=\pi$ is the only interval where \emph{none} of the hyperfine and fine resonances is effective.

While only binaries that nearly counter-rotate get through hyperfine and fine resonances without destroying the cloud, they might not be the only ones that carry it up to the Bohr region. The separations where these resonances are located can be even larger than the scale where binary formation mechanisms or other astrophysical interactions take place \cite{Amaro-Seoane:2012lgq,LISA:2022yao,LISA:2022yao,Stone:2016wzz,Bartos:2016dgn,Mckernan:2017ssq,Levin:2006uc}. We do not dive into further details here and simply consider close binary formation as a mechanism to ``skip'' early resonances.

\begin{figure}
       \centering
       \includegraphics[width=0.98\columnwidth,trim={0pt 12pt 0pt 0}]{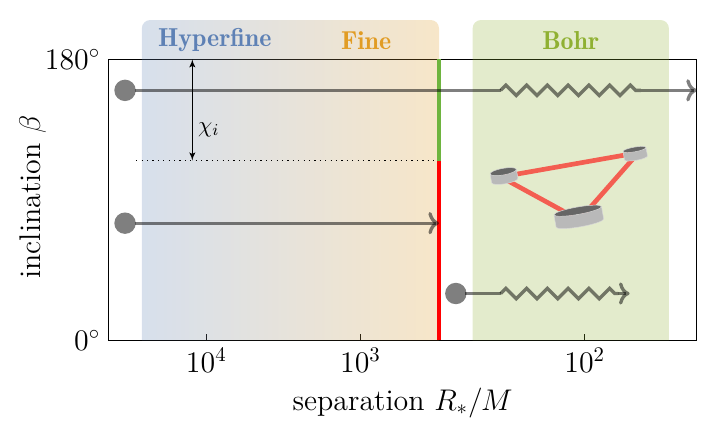}
       \caption{Schematic summary of the resonant history of the cloud-binary system. When the inclination angle falls inside an angular interval $\chi_i$ around a counter-rotating configuration, the cloud survives all the resonances (green line), becoming observable late in the inspiral. Otherwise, the cloud is destroyed (red line), leaving a distinctive mark on the orbital parameters. Binaries that form at small radii are an exception: they may skip the destructive (hyper)fine resonances.}
       \label{fig:schematic-illustration} 
 \end{figure}

The plethora of Bohr resonances encountered late in the inspiral, responsible for direct observational signatures, are almost all weak and sinking, meaning that they too do not change the state of the cloud. The only possible exception among the initial states we considered is $\ket{322}\to\ket{211}$. These can float and populate a relatively long-lived final state; however, the cloud is very efficiently ionized during the float and, thus, loses most of its mass. A summary of the possible resonant histories is given in Fig.~\ref{fig:schematic-illustration}.

\vskip 4pt
\paragraph*{\bf Direct observational signatures.} Understanding the resonant history of the system allows us to narrow down its impact on the waveform. The cloud either is destroyed or remains in its original state while the binary is nearly counter-rotating. The possible shapes of the ionization-driven frequency chirp \cite{Baumann:2022pkl} are then simply the ones induced by the state initially populated by superradiance, most likely $\ket{211}$ or $\ket{322}$, with $\beta\approx\pi$.

The sequence of sinking Bohr resonances encountered by the system \cite{Tomaselli:2024bdd} is also completely fixed by the initial state of the cloud. Their frequency \cite{Baumann:2018vus},
\beq
f_{\scalebox{0.60}{$\mathrm{GW}$}}^{\rm res}=\frac{\text{26}\,\text{mHz}}g\biggl(\frac{10^4M_\odot}{M}\biggr)\biggl(\frac\alpha{0.2}\biggr)^3\biggl(\frac1{n_a^2}-\frac1{n_b^2}\biggr)\,,
\label{eqn:position-resonances}
\eeq
can fall in the LISA band.\footnote{For $n_b\rightarrow \infty$, one recovers from \eqref{eqn:position-resonances} the position of the ionization kinks \cite{Baumann:2021fkf,Baumann:2022pkl,Tomaselli:2023ysb}.} Our work predicts the \emph{amplitude} of the corresponding frequency ``jump'',
\beq
\begin{aligned}
\Delta f_{\scalebox{0.60}{$\mathrm{GW}$}}&=\frac{\text{0.61}\,\text{mHz}}{\Delta m^{1/3}}\,\biggl(\frac{10^4M_\odot}{M}\biggr)\biggl(\frac{M_{\rm c}/M}{10^{-2}}\biggr)\biggl(\frac{q}{10^{-3}}\biggr)^{-1}\\&\times\biggl(\frac\alpha{0.2}\biggr)^3\biggl(\frac1{n_a^2}-\frac1{n_b^2}\biggr)^{4/3}\biggl(\frac{\abs{c_b}^2}{10^{-3}}\biggr)\,,
\label{eqn:jump-resonances}
\end{aligned}
\eeq
where $|c_{b}|^{2}$ can be determined through \eqref{eqn:scaling-cb2} and is typically $\sim\mathcal O(10^{-3})$. The dephasing induced by even a single jump can be $\sim\mathcal O(10^4)$ radians, which is well above the expected LISA precision.

\begin{figure*}
\centering
\includegraphics[]{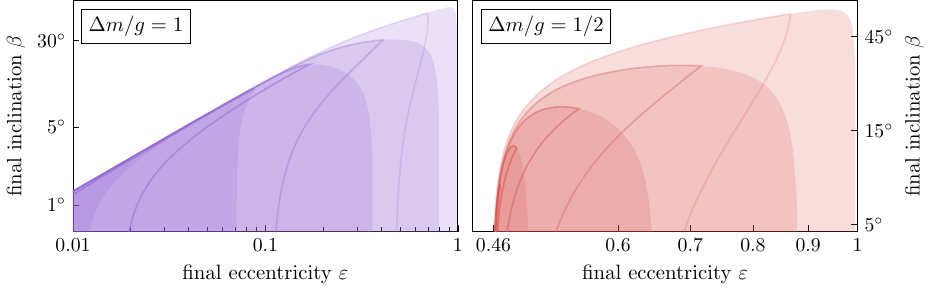}
\caption{The shaded regions show the possible values of eccentricity $\varepsilon$ and inclination $\beta$ at the completion of a floating resonance, starting from any initial values $\varepsilon_0$ and $\beta_0$, for different values of $D$. The values used, from the outermost to the innermost region, are $D=1,1.5,2,2.5,3$ (\emph{left panel}) and $D=1.8,2.6,3.4,4.2,5$ (\emph{right panel}); cf.~\eqref{eqn:D}. When the initial inclination is required to satisfy the conditions necessary to sustain the float, a smaller portion of each region is reachable. We enclosed in solid lines the reachable portions for $\beta_0\le\SI{128}{\degree}$ (\emph{left panel}) and $\beta_0\le\SI{142}{\degree}$ (\emph{right panel}), which are the thresholds for $\ket{211}\to\ket{21\,\minus1}$ and $\ket{211}\to\ket{210}$, for the reference parameters used in \eqref{eqn:chi_1}.}
\label{fig:parameter-range} 
\end{figure*}

\vskip 4pt
\paragraph*{\bf Indirect observational signatures.} Perhaps our most striking result is that, even when the cloud is destroyed during an early resonance, it still leaves a permanent, detectable mark on the binary. It does so by severely affecting the eccentricity $\varepsilon$ and inclination $\beta$ during the floating orbit, as shown in Fig.~\ref{fig:streamplot}. This mechanism depends on two parameters: the ratio $\Delta m/g$, which determines the fixed point $\varepsilon$ tends to (see eq.~\eqref{eqn:fixedpoints}), and
\beq
D=D_0\biggl(\frac{-g}2\biggr)^{\!2/3}\biggl(\frac{M_{\rm c}/M}{10^{-2}}\biggr)\biggl(\frac{q}{10^{-3}}\biggr)^{\!-1}\biggl(\frac\alpha{0.2}\biggr)\biggl(\frac{\tilde a}{0.5}\biggr)^{\!1/3}\,,
\label{eqn:D}
\eeq
which determines how closely the binary approaches it. For the two strongest hyperfine resonances from $\ket{211}$ ($\ket{322}$), the parameter $D_0$ assumes the values $3.30$ and $4.16$ ($1.28$ and $1.62$) respectively.

We show in Fig.~\ref{fig:parameter-range} the possible values of $\varepsilon$ and $\beta$ at the end of a floating resonance, as functions of $D$ and for two values of $\Delta m/g$. The float brings the orbit significantly close to the equatorial plane, even for large initial inclinations. An abundance of quasi-planar inspiral events can thus be indirect evidence for boson clouds. Whether the formation mechanisms of the binary, or other astrophysical processes, also lead to a natural preference for small inclinations is still subject to large uncertainties \cite{Amaro-Seoane:2012lgq,Pan:2021ksp,Pan:2021oob}.

Additionally, the eccentricity is suppressed by the main tones ($g=\Delta m$) and brought close to, or above, a nonzero fixed point by overtones ($g>\Delta m$). The latter scenario is especially interesting for binaries that are not dynamically captured, such as in the case of comparable mass ratios, because they are generally expected to be on quasi-circular orbits. The past interaction with a cloud can overturn this prediction. The float-induced high eccentricities are mitigated by the subsequent GW emission, but the binary will remain more eccentric than it would have been otherwise, even in late stages of the inspiral.

\vskip 4pt
\paragraph*{\bf Conclusions.} In this \emph{Letter}, we answered one of the outstanding questions about the phenomenology of gravitational atoms in binary systems. By studying their resonant history, we are able to show that either the cloud remains in its original state until late stages of the inspiral, or it is destroyed during an early floating resonance. In the first case, the binary is expected to be nearly counter-rotating with the cloud, which pinpoints uniquely its direct observational signatures given by Bohr resonances and ionization.

In the second case, the binary's eccentricity and inclination evolve toward a fixed point. The possibility of inferring the past existence of a cloud from its \emph{legacy} left on the binary parameters is a new and exciting observational prospect for both ground-based GW detectors, such as LIGO-Virgo or the Einstein Telescope, and space-borne ones, such as LISA. Proper population studies will be needed to turn this prediction into a test of fundamental physics with the available and future data.

\vskip 4pt
\paragraph*{\it Acknowledgements} We thank Rafael Porto for useful discussions and for sharing a draft of their related work \cite{Boskovic:2024fga}. TS is supported by the VILLUM Foundation (grant no.~VIL37766), the Danish Research Foundation (grant no.~DNRF162), and the European Union’s H2020 ERC Advanced Grant ``Black holes: gravitational engines of discovery'' grant agreement no.~Gravitas–101052587.

\clearpage
\bibliography{main}

\end{document}